\documentclass[sigconf, twocolumn]{acmart} 
\pdfoutput=1
\AtBeginDocument{%
  \providecommand\BibTeX{{%
    \normalfont B\kern-0.5em{\scshape i\kern-0.25em b}\kern-0.8em\TeX}}}
\copyrightyear{2022} 
\acmYear{2022} 
\acmConference[Issac' 22]{ProceeThe International Symposium on Symbolic and Algebraic Computationdings of the 2022 International Symposium on Symbolic and Algebraic Computation}
\acmBooktitle{The International Symposium on Symbolic and Algebraic Computation}
\acmYear{2022} 
\setcopyright{acmcopyright}

\usepackage[utf8]{inputenc}

\usepackage{graphicx}
\usepackage{pgf,tikz,pgfplots}
\pgfplotsset{compat=1.15}
\usetikzlibrary{arrows}
\usepackage{fancyhdr}
\usepackage{multirow}
\usepackage{enumerate}

\usepackage{math_defination}

\usepackage[ruled,vlined,linesnumbered]{algorithm2e}

\begin{document}

\title{Square-free Strong Triangular Decomposition of Zero-dimensional Polynomial Systems}
\author{Haokun Li, Bican Xia, Tianqi Zhao}
\email{haokunli, zhaotq@pku.edu.cn} 
\email{xbc@math.pku.edu.cn}
\affiliation{%
  \institution{School of Mathematical Sciences, Peking University}
  \state{Beijing}
  \country{China}}

\begin{abstract}
Triangular decomposition with different properties has been used for various types of problem solving, e.g. geometry theorem proving, real solution isolation of zero-dimensional polynomial systems, etc.
In this paper, the concepts of strong chain and square-free strong triangular decomposition (SFSTD) of zero-dimensional polynomial systems are defined. Because of its good properties, SFSTD may be a key way to many problems related to zero-dimensional polynomial systems, such as real solution isolation and computing radicals of  zero-dimensional ideals. 
Inspired by the work of Wang and of Dong and Mou, we propose an algorithm for computing SFSTD based on Gr\"obner bases computation.
The novelty of the algorithm is that we make use of saturated ideals and separant to ensure that the zero sets of any two strong chains have no intersection and every strong chain is square-free, respectively.
On one hand, we prove that the arithmetic complexity of the new algorithm can be  single exponential in the square of the number of variables, which seems to be among the rare complexity analysis results for triangular-decomposition methods.
On the other hand, we show experimentally that, on a large number of examples in the literature, the new algorithm is far more efficient than a popular triangular-decomposition method based on pseudo-division.
Furthermore, it is also shown that, on those examples, the methods based on SFSTD for real solution isolation and for computing radicals of zero-dimensional ideals are very efficient. 
\end{abstract}

\ccsdesc[500]{Computing methodologies~Algebraic algorithms}

\keywords{Triangular decomposition, Zero-dimensional polynomial system, Gr\"obner basis, Strong chain, Square-free strong chain}

\maketitle

\vspace{-4mm}
\section{Introduction}\label{sec:introduction}
\vspace{-0.5mm}

Decomposing a polynomial system to finitely many triangular sets with corresponding zero decomposition, called triangular decomposition of polynomial systems, is one of the fundamental tools in computational ideal theory. Wu first proposed a triangular-decomposition algorithm for computing characteristic sets in \cite{tun1978decision}, which was applied to geometry theorem proving. Since then, triangular decomposition methods have been applied successfully to not only geometry theorem proving but also lots of problems with diverse backgrounds, such as automated reasoning, real solution isolation, real solution classification and computing the radical of a polynomial ideal \cite{wenjun1984basic,chou1990ritt,tun1978decision,yang1990searching,yang1992criterion,xia2002algorithm,xia2006real,wu2006automated,boulier2014real,gao1992introduction,cheng2012root,xia2016automated,chen2015generic}, to name a few.
Different applications may require different types of triangular sets and triangular decomposition with specific properties (see the book \cite{DBLP:series/tmsc/Wang01} for reference).
In the passed several decades, many specific triangular sets or triangular systems have been defined, e.g. regular chains, normal chains and square-free triangular sets. 
And, lots of triangular-decomposition algorithms have been proposed, see for example \cite{yang1990searching, aubry1999theories,  kalkbrener1993generalized, lazard1992solving, wang2000zero, wang1996solving, DBLP:conf/casc/DongM19, wang2000computing, dong2017decomposing, chen2011solving, hubert2001notes, chen2012algorithms}.
Nevertheless, classical triangular-decomposition algorithms are mainly based on factorization and pseudo-division and are well known not so efficient on big examples. Another significant but somehow neglected aspect is that, there are few results about the complexity of classical triangular-decomposition algorithms. 

The Gr\"obner basis method, first proposed by Buchberger in \cite{buchberger1965algorithmus}, has been extensively studied and applied to lots of research fields. It is well known that the complexity of computing Gr\"obner bases can be double-exponential time in general case and  single-exponential time for zero-dimensional ideals \cite{hashemi2005complexity,lakshman1991single}.
There are some famous algorithms for computing Gr\"obner bases (see for example \cite{faugere1999new, faugere2002new, DBLP:journals/iacr/GaoVW10}) and corresponding tools are available in some computer algebra systems, e.g. {\tt Maple}, {\tt Mathematica} and {\tt Magma}.



Then, one may ask whether there exists a connection between triangular sets and Gr\"obner bases and whether one can obtain triangular decomposition by Gr\"obner bases computation.
The problem was well solved in Wang's work \cite{DBLP:journals/mics/Wang16} in 2016.
The key concept is the W-characteristic set, which is a minimal triangular set extracted from a reduced Gr\"obner basis with respect to (w.r.t.) the lex ordering (written as reduced LEX Gr\"obner basis).
Wang proved that if the variable ordering condition is satisfied for a W-characteristic set, then the regularity and normality of the W-characteristic set are equivalent.
Later, Dong and Mou proposed some algorithms in \cite{dong2017decomposing, DBLP:conf/casc/DongM19} for characteristic decomposition which is also a type of triangular decomposition consisting of normal chains.
Owe to efficient computation of Gr\"obner bases,
their algorithms perform better on some complicated polynomial systems than the classical algorithms.

In this paper, we only consider zero-dimensional systems.
We define a new type of triangular sets, namely {\em strong chains} (see Def. \ref{def:PureChain}). And then, a new type of triangular decomposition, called square-free strong triangular decomposition (SFSTD) (see Def. \ref{def:sfstd}), is introduced for real solution isolation.
We observe that SFSTD can also be applied to computing radicals of zero-dimensional ideals.
So, the main goal of this paper is to efficiently compute SFSTD of zero-dimensional systems.
More formally, we have the following problem statement:\\
{\bf Input:} A nonempty finite set $F\subseteq\Q[x_1,\ldots,x_n]\setminus\{0\}$.\\
{\bf Output:} An SFSTD $\{\mT_1,\ldots,\mT_s\}$ of $F$, where each $\mT_i$ is a square-free strong chain, such that
\vspace{-1mm}
\begin{itemize}
    \item the zero set of $F$ equals the union of all zero sets of $\mT_i$, and
    \item any two zero sets of $\mT_i$ and $\mT_j$ $(i\ne j)$ have no intersection,
\end{itemize}
\vspace{-1mm}
if $F$ is zero-dimensional; FAIL otherwise.\\
We list our main contributions as follows:
\vspace{-1mm}
\begin{enumerate}[1.]
    \item Based on \cite[Algorithm 1]{DBLP:conf/casc/DongM19}, we propose an algorithm (Algorithm \ref{alg:StrongTriDec}) for computing a type of triangular decomposition consisting of strong chains.
    \item We propose an algorithm (Algorithm \ref{alg:StrongSfTriDec}) for computing SFSTD.
    \item We prove that the arithmetic complexity of Algorithm \ref{alg:StrongSfTriDec} can be  single exponential in the square of the number of variables.
    \item We implemented Algorithm \ref{alg:StrongSfTriDec} with {\tt Maple2021}. Our experiments show that Algorithm \ref{alg:StrongSfTriDec} is far more efficient than the classical method for zs-rc decomposition in \cite{boulier2014real} (see the group of columns TD in Table \ref{tab:result}).
    \item By SFSTD, one can compute isolating cubes of real solutions to the system, and the radical of the ideal generated by the system, efficiently (see the groups of columns RSI and RA in Table \ref{tab:result}).
\end{enumerate}
\vspace{-1mm}
The rest of this paper is organized as follows.
In Section \ref{sec:preliminary}, we recall some basic concepts and existing results about the theories of Gr\"obner bases and triangular sets.
In Section \ref{sec:strongtridec}, we propose Algorithm \ref{alg:StrongTriDec}.
The termination and correctness of Algorithm \ref{alg:StrongTriDec} is guaranteed by Theorem \ref{thm:algStrongTriDecIsCorrect}.
In Section \ref{sec:strongsftridec}, we propose Algorithm \ref{alg:StrongSfTriDec} and prove the termination and correctness 
(see Theorem \ref{thm:algStrongSfTriDecIsCorrect}).
In Section \ref{sec:complexity}, we analyze the arithmetic complexity of Algorithm \ref{alg:StrongSfTriDec}.
In Section \ref{sec:applications}, we present two applications of SFSTD: real solution isolation and computing the radical of a zero-dimensional ideal.
In Section \ref{sec:experiments}, we explain the implementation details and show the experimental results. Section \ref{sec:conclusion} concludes the paper.

\vspace{-2.5mm}
\section{Preliminary}\label{sec:preliminary}
\vspace{-0.5mm}

In the section, we recall some basic concepts and existing results about the theories of Gr\"obner bases and triangular sets.
The reader is referred to \cite{DBLP:books/daglib/0071992, DBLP:series/tmsc/Wang01} for more details.

\vspace{-2.5mm}
\subsection{Zero-dimensional Systems}\label{subsec:GB-TS-TD}
\vspace{-0.5mm}

Let $x_1,\ldots,x_n$ be $n$ variables and let $\vX$ denote the vector $(x_1,\ldots,x_n)$.
Throughout the paper, we fix the variable ordering $x_1<\cdots<x_n$.

$\Q$ denotes the rational numbers and $\CC$ denotes the complex numbers. 
Let $F$ be any polynomial set in $\Q[\vX]$.
We denote by $\langle F\rangle$ \emph{the ideal} generated by $F$ in $\CC[\vX]$.
For any ideal $I\subseteq\CC[\vX]$, $\V(I)$ denotes the {\em affine variety} $\{(a_1,\ldots,a_n)\in\CC\mid f(a_1,\ldots,a_n)=0~{\rm for~all}~f\in I\}$.
In particular, $\V(F):=\V(\langle F\rangle)$.
For any polynomial $f\in\Q[\vX]$ with an admissible monomial ordering, $\lt(f)$ denotes \emph{the leading term} of $f$ and $\lm(f)$ denotes \emph{the leading monomial} of $f$. 

\vspace{-1.5mm}
\begin{definition}
For $F\subseteq\Q[\vX]$, $F$ is  a \emph{zero-dimensional system} or $\langle F\rangle$ is  a \emph{zero-dimensional ideal}, if $\V(F)$ is a finite set.
\end{definition}
\vspace{-2mm}
\vspace{-2mm}

\begin{definition}
Fix a monomial ordering and let $F\subseteq\Q[\vX]$. 
A finite subset $G=\{g_1,\ldots,g_t\}\subseteq\Q[\vX]$ is called a \emph{Gr\"obner basis} of $\langle F\rangle$ if $\langle\lt(g_1),\ldots,\lt(g_t)\rangle=\langle\lt(\langle F\rangle)\rangle$, where $\lt(\langle F\rangle)=\{\lt(f)\mid f\in \langle F\rangle\}$.
\end{definition}

\vspace{-2mm}


\vspace{-2mm}
\begin{proposition}
[{\cite[Chap. 8.3]{DBLP:books/daglib/0071992}}]
\label{prop:ZeroDimEquivCon}
Let $F\subseteq\Q[x_1,\ldots,x_n]$.
The following statements are equivalent:
\begin{enumerate}[(a)]
    \item $F$ is a zero-dimensional system,
    \item for every Gr\"obner basis $G$ of $\langle F\rangle$, $G$ contains $n$ polynomials $g_1,\ldots,\allowbreak g_n$ such that $\lm(g_i)=x_i^{k_i}$ ($k_i\ge1$),
    \item there exist a monomial ordering $\prec$ and a Gr\"obner basis of $\langle F\rangle$ with respect to $\prec$ which contains $n$ polynomials $g_1,\ldots,g_n$ such that $\lm(g_i)=x_i^{k_i}$ ($k_i\ge1$).
\end{enumerate}
\end{proposition}
\vspace{-2mm}

\vspace{-2.5mm}
\subsection{Triangular Sets}
\vspace{-0.5mm}

Let $f\in\Q[\vX]\setminus\Q$.
We denote by $\lv(f)$ the \emph{main variable} of $f$ and
by $\ini(f)$ the \emph{initial} (or leading coefficient w.r.t. $\lv(f)$) of $f$.
For any $F\subseteq\Q[\vX]$, $\lv(F):=\{\lv(f)\mid f\in F\}$ and $\ini(F):=\{\ini(f)\mid f\in F\}$.
Let $\mT=[T_1,\ldots,T_t]$ be a finite nonempty list of nonconstant polynomials in $\Q[x_1,\ldots,x_n]$.
$\mT$ is called a \emph{triangular set} if
$\lv(T_1)<\cdots<\lv(T_t)$. 


Let $f$ and $g$ be two polynomials in $\Q[\vX]$, $F\subseteq\Q[\vX]$ and $\mT=[T_1,\ldots,T_t]\subseteq\Q[\vX]$ be a triangular set.
We denote by $\deg(f,x_i)$ the {\em degree} of $f$ w.r.t. a particular variable $x_i$ and by $\sep(f)$ the \emph{separant} of $f$, i.e. $\partial f/\partial\lv(f)$.
The \emph{saturated ideal} of $F$ w.r.t. $f$ is defined as $\langle F\rangle:f^{\infty}:=\{g\in\CC[\vX]\mid {\rm there~exists}~i\ge0~{\rm such~that}~f^{i}g\in\langle F\rangle\}$.
And, we denote by $\sat(\mT)$ the saturated ideal of $\mT$, namely $\langle\mT\rangle:(\ini(T_1)\cdots\ini(T_t))^{\infty}$.
The \emph{resultant} of $f$ and $g$ w.r.t. $\lv(g)$ is denoted by $\res(f,g)$, and the resultant of $f$ and $\mT$ is defined as $\res(f,\mT):=\res(\cdots\res(\res(f,T_t),T_{t-1}),\ldots,T_1)$.
$\mT$ is called a \emph{regular chain} (or is said to be \emph{regular}) if $\ini(T_1)\ne0$ and for each $i$ ($2\le i\le t$), $\res(\ini(T_i),[T_1,\ldots,T_{i-1}])\ne0$.
$\mT$ is called a \emph{normal chain} (or is said to be \emph{normal}) if $\ini(\mT)$ does not involve the main variables of $\mT$.
It is clear that any normal chain is a regular chain.
$\mT$ is said to be \emph{square-free} if 
the discriminant of $T_1$ w.r.t. $\lv(T_1)$ is not equal to $0$, and for each $i$ ($2\le i\le t$), $\res(\sep(T_i),[T_1,\ldots,T_{i}])\ne0$.

\vspace{-2.5mm}
\subsection{W-Characteristic Sets}
\vspace{-0.5mm}


\begin{definition}[\cite{DBLP:journals/mics/Wang16}]\label{def:W-set}
Let $F\subseteq\Q[\vX]$ and $G\subseteq\Q[\vX]$ be the reduced Gr\"obner basis of $\langle F\rangle$ w.r.t. the lex ordering $\preclex$. 
Define $G_i:=\{g\in G\mid\lv(g)=x_i\}$ for $i=1,\ldots,n$. 
For every nonempty $G_i$, let $g_i\in G_i$ be the polynomial such that $\lm(g_i)\preclex\lm(g)$ for any $g\in G_i\setminus\{g_i\}$.  
The ordered list of all $g_i$ 
is called the {\em W-characteristic set} of $F$.
\end{definition}

\vspace{-2mm}


For the variable ordering $x_1<\cdots<x_n$,
we say that \emph{the variable ordering condition} is satisfied for a W-characteristic set $\mC$,
if the variables in $\{x_1,\ldots,x_n\}\setminus\lv(\mC)$ are ordered before $\lv(\mC)$. 
The following theorem presents two properties of W-characteristic sets.

\vspace{-1.5mm}

\begin{theorem}
[\cite{DBLP:journals/mics/Wang16}]
\label{prop:WcharProperty}
Let $\mC=[C_1,\ldots,C_t]$ be the W-characteristic set of $F\subseteq\Q[x_1,\ldots,x_n]$, where $1\le t\le n$. 
We have: 
\begin{enumerate}[(a)]
    \item \label{item:pro:WcharProperty2}$\langle\mC\rangle\subseteq\langle F\rangle\subseteq\sat(\mC)$, and
    \item \label{item:pro:WcharProperty3}if the variable ordering condition is satisfied for $\mC$ and $\mC$ is not a normal chain, then there exists an integer $k~(1\le k< t)$ such that $[C_1,\ldots,C_k]$ is normal and $[C_1,\ldots,C_{k+1}]$ is not regular. 
\end{enumerate}
\end{theorem}
\vspace{-1.5mm}

\vspace{-2.5mm}
\section{STD}\label{sec:strongtridec}
\vspace{-0.5mm}

In order to compute SFSTD, we need to compute a type of triangular decomposition consisting of strong chains first.
And, the algorithm \cite[Algorithm 1]{DBLP:conf/casc/DongM19} of Dong and Mou can calculate such decomposition for zero-dimensional systems.
Since our purpose is real solution isolation, 
we also require the zero sets of the strong chains in the decomposition are pairwise disjoint.
We call the required decomposition {\em strong triangular decomposition} (STD). 
Section \ref{subsec:DefSTD} presents the formal definitions of strong chain and STD, and
Section \ref{subsec:AlgSTD} presents an algorithm for computing STD which is based on \cite[Algorithm 1]{DBLP:conf/casc/DongM19}.

\vspace{-2.5mm}
\subsection{The Definition of STD}\label{subsec:DefSTD}
\vspace{-0.5mm}


\begin{definition}\label{def:PureChain}
Let $\mT=[T_1,\ldots,T_n]\subseteq\Q[x_1,\ldots,x_n]$ be a triangular set.
$\mT$ is called a \emph{strong chain} (or is said to be \emph{strong}) if $\ini(\mT)\subseteq\Q\setminus\{0\}$ and $\lv(T_i)=x_i$ for $i=1,\ldots,n$.
A strong chain is said to be \emph{reduced}, if for each $i$ $(1\le i\le n)$, 
$\ini(T_i)=1$ and $\deg(T_i,x_i)>\deg(T_j,x_i)$ where $j>i$.
\end{definition}
\vspace{-1.5mm}
Any strong chain is a normal chain.
And, strong chains have very good properties.
\vspace{-1.5mm}
\begin{proposition}\label{prop:PureChainIsGro}
Fix the lex monomial ordering and let $\mT=[T_1,\ldots,T_n]\subseteq\Q[\vX]$ be a strong chain.
Then,
\begin{enumerate}[(a)]
    \item $\sat(\mT)=\langle\mT\rangle$, \label{item:PureChainIsGro1}
    \item $\lm(T_i)=x_{i}^{j_i}\;(j_i\ge1)$ for $i=1,\ldots,n$, \label{item:PureChainIsGro2}
    \item $\mT$ is a LEX Gr\"obner basis, and\label{item:PureChainIsGro3}
    \item $\mT$ is a zero-dimensional system.\label{item:PureChainIsGro4}
\end{enumerate}
Further, $\mT$ is reduced if and only if $\mT$ is a reduced LEX Gr\"obner basis. 
\end{proposition}
\vspace{-3mm}
\begin{proof}
(\ref{item:PureChainIsGro1}) It is because $\ini(\mT)\subseteq\Q\setminus\{0\}$.
(\ref{item:PureChainIsGro2}) It is clear.
(\ref{item:PureChainIsGro3}) By (\ref{item:PureChainIsGro2}), for any $i_1\ne i_2$, the greatest common divisor of $\lm(T_{i_1})$ and $\lm(T_{i_2})$ is $1$.
Then, by \cite[Lemma 5.66]{DBLP:books/daglib/0071992}, we complete the proof.
(\ref{item:PureChainIsGro4}) It is obvious by (\ref{item:PureChainIsGro2}),  (\ref{item:PureChainIsGro3}) and  Proposition \ref{prop:ZeroDimEquivCon}.
\end{proof}
\vspace{-1.5mm}
\vspace{-1.5mm}
\begin{definition}
Let $F\subseteq\Q[\vX]$ be a zero-dimensional system. 
A \emph{strong triangular decomposition} (STD) of $F$ is a finite set of strong chains $\{\mT_1,\ldots,\mT_s\}\subseteq\Q[\vX]$ such that 
\begin{align*}\label{eq:def-StrongTriDec}
      \V(F)~=~\bigcup_{i=1}^{s}\V(\mT_i)~{\rm and}~\V(\mT_i)\cap\V(\mT_j)=\emptyset~{\rm for~any~}i\ne j.
\end{align*}
\end{definition}




\vspace{-1.5mm}
\vspace{-2.5mm}
\subsection{Computing STD}\label{subsec:AlgSTD}
\vspace{-0.5mm}
\subsubsection{The Algorithm}
\quad \par

Given a nonempty finite polynomial set $F\subseteq\Q[\vX]\setminus\{0\}$, if $F$ is zero-dimensional, then Algorithm \ref{alg:StrongTriDec} computes an STD of $F$.
(Note that when $\V(F)=\emptyset$, the output is an empty set.)
Otherwise, the output is FAIL.
The process of Algorithm \ref{alg:StrongTriDec} is as follows.

Let $\Phi$ be a set of polynomial sets for STD (initialized as $\{F\}$), and $ans$ be a set of computed strong chains (initialized as $\emptyset$).
In the while loop, 
for the first time, we pick $F$ and compute the reduced LEX Gr\"obner basis $G$ of $\langle F\rangle$.
If there exists $x_i$ such that for any $g\in G$, $\lm(g)\ne x_i^{k}$ ($k\ge1$), then $F$ is not zero-dimensional. 
Otherwise, $F$ is zero-dimensional. 
We extract the W-characteristic set $\mC$ from $G$.
Suppose that $\mC=[C_1,\ldots,C_m]$, where $m\le n$.
\begin{enumerate}[I.]
    \item If $\mC$ is a strong chain, then we add $\mC$ to $ans$.  \label{item:AlgSTD1}
    \item \label{item:AlgSTD2} If $\mC$ is not strong, then we compute $k$ which is the smallest integer that makes $[C_1,\ldots,C_k]$ not strong. 
    Let $G_{sat}$ be the reduced Gr\"obner basis of $\langle  C_1,\ldots,C_{k-1}\rangle:\ini(C_k)^{\infty}$ w.r.t. any monomial ordering. We update $\Phi$ with $G\cup\{\ini(C_k)\}$ and $G\cup G_{sat}$.
\end{enumerate}

For the $i$-th time ($i\ge2$), we pick $P$ from $\Phi$ and compute the W-characteristic set $\mC$ of $P$.
Then, repeat (\ref{item:AlgSTD1}) and (\ref{item:AlgSTD2}) in the above paragraph.


\begin{algorithm}[t]
\scriptsize
\DontPrintSemicolon
\LinesNumbered
\SetKwInOut{Input}{Input}
\SetKwInOut{Output}{Output}
\Input{a nonempty finite set $F\subseteq\Q[\vX]\setminus\{0\}$ and the vector $\vX$}
\Output{$ans=\{\mT_1,\ldots,\mT_s\}$, a finite set of strong chains such that
\[\V(F)~=~\bigcup_{\mT_i\in ans}\V(\mT_i)~{\rm and}~\V(\mT_i)\cap\V(\mT_j)=\emptyset~{\rm for~any~}i\ne j,
\]
if $F$ is zero-dimensional;
FAIL otherwise}
\caption{{\bf STD}}\label{alg:StrongTriDec}
\BlankLine
$ans\leftarrow\emptyset,~\Phi\leftarrow\{F\},~num\leftarrow0$\;
\While{$\Phi\ne\emptyset$}{
$num\leftarrow num+1$\;
Choose $P$ from $\Phi$ and set $\Phi\leftarrow\Phi\setminus\{P\}$\; \label{algline:choosep}
$G\leftarrow$ the reduced LEX Gr\"obner basis of $\langle P\rangle$\label{line:StrongTriDec-ReGB}\;
\If{$G\ne\{1\}$\label{line:StrongTriDec-1}}
{\If{$num=1$ and 
there exists $x_i$ such that for any $g\in G$, $\lm(g)\ne x_i^{k}$ ($k\ge1$)}
 {\Return{\rm FAIL}\label{line:StrongTriDec-FAIL}}
 {$\mC\leftarrow[C_1,\ldots,C_m]~(m\le n)$ which is the W-characteristic set of $G$\;
 \# In fact, we have $m=n$ (see the termination proof of Theorem \ref{thm:algStrongTriDecIsCorrect}).\;
 \eIf{$\mC$ is a strong chain\label{line:StrongTriDec-strong}}
 {$ans\leftarrow ans\cup\{\mC\}$\label{line:StrongTriDec-pure}}
 {
 $C_k\leftarrow$ the first polynomial of $\mC$ that makes $[C_1,\ldots,C_k]$ not strong\;
 $G_{sat}\leftarrow$ the reduced Gr\"obner basis of $\langle C_1,\ldots,C_{k-1}\rangle:\ini(C_k)^{\infty}$ w.r.t. any monomial ordering \label{line:StrongTriDec-Gsat}\;
 $\Phi\leftarrow\Phi\cup\{G\cup\{\ini(C_k)\}\}\cup\{G\cup G_{sat}\}$\label{line:StrongTriDec-Spilt}
 }
 }
}}
\Return{$ans$}
\end{algorithm}


Algorithm \ref{alg:StrongTriDec} is illustrated on an example.
\vspace{-1.5mm}
\begin{example}
Consider $F=\{x^2-1,xy-y,y^3-y\}\subseteq\Q[x,y]$ with $x<y$.
It is clear that $F$ itself is a reduced LEX Gr\"obner basis.
And, by Proposition \ref{prop:ZeroDimEquivCon}, $F$ is zero-dimensional.
The W-characteristic set of $F$ is $[x^2-1,xy-y]$.
Note that it is not a strong chain and
$xy-y$ is the first polynomial that makes it not strong.
Then, $\Phi$ is updated with two elements $F\cup\{x-1\}$ and $F\cup\{x+1\}$ in Line \ref{line:StrongTriDec-Spilt}.
For the element $F\cup\{x-1\}$,
the reduced LEX Gr\"obner basis of $\langle F\cup\{x-1\}\rangle$ is $\{x-1,y^3-y\}$, which is also the W-characteristic set.
Since $[x-1,y^3-y]$ is a strong chain,
we add it to the set of computed strong chains.
For the element $F\cup\{x+1\}$,
from the reduced LEX Gr\"obner basis $\{x+1,y\}$ of $\langle F\cup\{x+1\}\rangle$,
the W-characteristic set is $[x+1,y]$.
Since $[x+1,y]$ is strong,
Algorithm \ref{alg:StrongTriDec} terminates with $\Phi=\emptyset$.
And, $\{[x-1,y^3-y],[x+1,y]\}$ is an STD of $F$.
\end{example}

\vspace{-3.5mm}

\subsubsection{Correctness and Termination of Algorithm \ref{alg:StrongTriDec}}\label{subsubsec:CorrectAlgStrongTriDec}
\quad \par
In order to prove the correctness and termination of Algorithm \ref{alg:StrongTriDec}, we prepare Lemma \ref{lemma:WcharIsStrong}, Proposition \ref{prop:ZeroSatIdeal} and Proposition \ref{prop:StrongChainOr}.

\vspace{-1.5mm}

\begin{lemma}\label{lemma:WcharIsStrong}
Let $\mC$ be the W-characteristic set of $F\subseteq\Q[\vX]$. 
If $\mC$ is a strong chain, then $\langle F\rangle=\langle\mC\rangle$.
\end{lemma}
\vspace{-3mm}
\begin{proof}
It is clear by Proposition \ref{prop:PureChainIsGro} (\ref{item:PureChainIsGro1}) and Theorem \ref{prop:WcharProperty} (\ref{item:pro:WcharProperty2}).
\end{proof}

\vspace{-1.5mm}
Let $\tau$ be a new variable.
Define $\pi$ as the \emph{canonical projection}: $\CC^{n+1}\rightarrow\CC^{n}$ such that
$\pi(x_1,\ldots,x_n,\tau)=(x_1,\ldots,x_n)$.
For any point set $W\subseteq\CC^{n}$, the smallest variety containing $W$ is denoted by $\overline{W}$.
For any affine variety $V\subseteq\CC^n$, define {\em the radical ideal} $\I(V):=\{f\in\CC[\vX]\mid f(a_1,\ldots,a_n)=0~{\rm for~any}~(a_1,\ldots,a_n)\in V\}$.

\vspace{-1.5mm}

\begin{proposition}\label{prop:ZeroSatIdeal}
Let $F\subseteq\Q[\vX]$ be a zero-dimensional system and $f\in\Q[\vX]$.
Then, $\V(\langle F\rangle:f^{\infty})=\V(F)\setminus\V(f)$.
\end{proposition}
\vspace{-3mm}

\begin{proof}
Note that
$\V(\langle F\rangle:f^{\infty})=\V(\langle F\cup\{1-\tau f\}\rangle\cap\CC[\vX])=\overline{\pi(\V(F\cup\{1-\tau f\}))}=\overline{\V(F)\setminus\V(f)} $.
Since $F$ is zero-dimensional, we have $\overline{\V(F)\setminus\V(f)}=\V(F)\setminus\V(f)$.
That completes the proof.
\end{proof}

\vspace{-3mm}

\begin{proposition}\label{prop:StrongChainOr}
Let $\mT\subseteq\Q[\vX]$ be a strong chain and $f\in\Q[\vX]$.
If $\res(f,\mT)=0$, 
there exists $g\in\CC[\vX]\setminus\langle\mT\rangle$ such that $fg\in\langle\mT\rangle$.
\end{proposition}
\vspace{-3mm}

\begin{proof}
Let $V_1:=\V(\mT)$ and $V_2:=\V(\mT)\setminus\V(f)$.
Since $\mT$ is a zero-dimensional system by Proposition \ref{prop:PureChainIsGro} (\ref{item:PureChainIsGro4}), $V_2$ is an affine variety.
By \cite[Theorem 2.2]{xia2016automated}, $\res(f,\mT)=0$ if and only if $\V(\mT)\cap\V(f)\ne\emptyset$, i.e., $\V_1\ne\emptyset$ and $V_2$ is a proper subset of $V_1$.
Then, it is equivalent to that there exists $q\in\I(V_2)\setminus\I(V_1)$.
Note that $fq\in\I(V_1)$.
Then, there exists some integer $k_1\ge1$ such that $(fq)^{k_1}\in \langle\mT\rangle$.
Let $k_2$ be the smallest integer number such that $f^{k_2}q^{k_1}\in\langle\mT\rangle$. 
Since $q\notin\I(V_1)$, $k_2\ge1$. 
Thus, we take $g=f^{k_2-1}q^{k_1}$ which is not in $\langle\mT\rangle$. 
\end{proof}

\vspace{-3mm}

\begin{theorem}\label{thm:algStrongTriDecIsCorrect}
Algorithm \ref{alg:StrongTriDec} terminates correctly.
\end{theorem}
\vspace{-3mm}

\begin{proof}
(Correctness) 
For the input polynomial set $F$, let $G_0$ be the reduced LEX Gr\"obner basis of $\langle F\rangle$.
By Proposition \ref{prop:ZeroDimEquivCon}, if there exists $x_i$ such that for any $g\in G_0$, $\lm(g)\ne x_i^{k}$ ($k\ge1$), then $F$ is not zero-dimensional. 
Algorithm \ref{alg:StrongTriDec} returns FAIL in Line \ref{line:StrongTriDec-FAIL}.

Otherwise, $F$ is zero-dimensional. 
In every loop, we pick a polynomial set $P$ from $\Phi$.
Let $G~(G\ne\{1\})$ be the reduced LEX Gr\"obner basis of $\langle P\rangle$ and $\mC=[C_1,\ldots,C_m]~(m\le n)$ be the W-characteristic set of $G$.
If $\mC$ is strong, then we add $\mC$ to the output set in Line \ref{line:StrongTriDec-pure}.
If $\mC$ is not strong, $\Phi$ is updated with two sets in Line \ref{line:StrongTriDec-Spilt}.
Therefore, we only need to prove that
\begin{enumerate}[(I)]
    \item \label{item:algStrongTriDecIsCorrect:Proof1} $\V(G)=\V(\mC)$, where $\mC$ is the strong W-characteristic set,
    \item \label{item:algStrongTriDecIsCorrect:Proof2}  $\V(G\cup\{\ini(C_k)\})\cap\V(G\cup G_{sat})=\emptyset$ and $\V(G)=\V(G\cup\{\ini(C_k)\})\cup\V(G\cup G_{sat})$, where $C_k$ is the first polynomial that makes $[C_1,\ldots,C_k]$ not strong and $G_{sat}$ is a Gr\"obner basis of $\langle C_1,\ldots,C_{k-1}\rangle:\ini(C_k)^{\infty}$.
\end{enumerate}
By Lemma \ref{lemma:WcharIsStrong}, (\ref{item:algStrongTriDecIsCorrect:Proof1}) is clear.
It remains to prove (\ref{item:algStrongTriDecIsCorrect:Proof2}).
Since $[C_1,\ldots,C_{k-1}]$ is strong, by Proposition \ref{prop:PureChainIsGro} (\ref{item:PureChainIsGro4}), $\{C_1,\ldots,\allowbreak C_{k-1}\}$ is zero-dimensional.
So, by Proposition \ref{prop:ZeroSatIdeal},
$\V(G\cup G_{sat})=\V(G)\cap(\V(C_1,\ldots,C_{k-1})\setminus\V(\ini(C_k))).$
Note that $\V(G)\subseteq\V(C_1,\ldots,C_{k-1})$. 
So, $\V(G\cup G_{sat})=\V(G)\setminus\V(\ini(C_k))$.
Then, because $\V(G\cup\{\ini(C_k)\})=\V(G)\cap\V(\ini(C_k))$, the proof is completed.

(Termination) 
If $F$ is not zero-dimensional, the termination is obvious.
Otherwise, the termination is equivalent to that every ideal generated by the added set in Line \ref{line:StrongTriDec-Spilt} is strictly larger than that generated by the removed one.
So, we only need to prove that $\langle G\cup\{\ini(C_k)\}\rangle$ and $\langle G\cup G_{sat}\rangle$ are both strictly larger than $\langle G\rangle$. Because $G$ is a reduced Gr\"obner basis, we have $\ini(C_k)\notin\langle G\rangle$.
Then, $\langle G\cup\{\ini(C_k)\}\rangle$ is strictly larger than $\langle G\rangle$. It remains to prove $\langle G\rangle\subsetneq\langle G\cup G_{sat}\rangle$.

Firstly, we prove that $\mC=[C_1,\ldots,C_n]~{\rm with}~\lv(C_i)=x_i$, i.e., $\lv(\mC)=\{x_1,\ldots,x_n\}$.
Since the input $F$ is zero-dimensional, by the proof of the correctness, every polynomial set in $\Phi$ is also zero-dimensional. 
Then, by Proposition \ref{prop:ZeroDimEquivCon} and Definition \ref{def:W-set}, we have $\lv(\mC)=\{x_1,\ldots,x_n\}$.

Secondly, we prove that $\res(\ini(C_{k}),[C_1,\ldots,C_{k-1}])=0$.
Note that $C_k$ is the first polynomial that makes $[C_1,\ldots,C_k]$ not strong.
Then,
$\mC$ and $[C_1,\ldots,C_{k}]$ are not normal, but $[C_1,\ldots,C_{k-1}]$ is normal.
Since $\lv(\mC)=\{x_1,\ldots,x_n\}$, the variable ordering condition is satisfied for $\mC$.
So, by Theorem \ref{prop:WcharProperty} (\ref{item:pro:WcharProperty3}), $[C_1,\ldots,C_{k}]$ is not regular.
Then, $\res(\ini(C_{k}),[C_1,\ldots,C_{k-1}])=0$.

Finally, we prove $\langle G\rangle\subsetneq\langle G\cup G_{sat}\rangle$, which is equivalent to proving that there exists 
$g\in\langle G_{sat}\rangle\setminus\langle G\rangle$.
Note that $[C_1,\ldots,C_{k-1}]$ is a strong chain in $\Q[x_1,\ldots,x_{k-1}]$.
So, by the conclusion in the above paragraph and by Proposition \ref{prop:StrongChainOr}, 
there exists $g\in\CC[x_1,\ldots,x_{k-1}]\allowbreak\setminus\langle C_1,\ldots,C_{k-1}\rangle$ such that $g\cdot\ini(C_k)\in\langle C_1,\ldots,C_{k-1}\rangle$.
Recall that $\langle G_{sat}\rangle=\langle C_1,\ldots,\allowbreak C_{k-1}\rangle:\ini(C_k)^{\infty}$.
So, we have $g\in\langle G_{sat}\rangle$.
Then, we only need to prove $g\notin\langle G\rangle$.
Note that $\langle G\rangle \cap\CC[x_1,\ldots,x_{k-1}]\allowbreak=\langle G  \cap\CC[x_1,\ldots,x_{k-1}]\rangle$ and $[C_1,\ldots,C_{k-1}]$ is the W-characteristic set of $G\cap\CC[x_1,\ldots,x_{k-1}]$. 
Thus, by Lemma \ref{lemma:WcharIsStrong},  $\langle G\rangle\cap\CC[x_1,\ldots,x_{k-1}]=\langle C_1,\ldots,C_{k-1}\rangle$.
Then, because $g\in\CC[x_1,\ldots,x_{k-1}]\allowbreak\setminus\langle C_1,\ldots,C_{k-1}\rangle$, $g\notin\langle G\rangle$. 
\end{proof}
\vspace{-3mm}
\begin{corollary}\label{coro:STD}
If $\{\mT_1,\ldots,\mT_s\}$ is an STD of a zero-dimensional system $F\subseteq\Q[\vX]$ computed by Algorithm \ref{alg:StrongTriDec}, then for each $i$, $\mT_i$ is reduced  and $\langle F\rangle\subseteq\langle\mT_i\rangle$.
\end{corollary}
\vspace{-3mm}
\begin{proof}
It is clear by the proof of Theorem \ref{thm:algStrongTriDecIsCorrect}, by Proposition \ref{prop:PureChainIsGro} and by Lemma \ref{lemma:WcharIsStrong}.  
\end{proof}
\vspace{-3mm}

\begin{remark}
Except for the following two aspects, 
Algorithm \ref{alg:StrongTriDec} is similar to \cite[Algorithm 1]{DBLP:conf/casc/DongM19}.
In order to guarantee the zero sets to be pairwise disjoint, Algorithm \ref{alg:StrongTriDec} computes $\langle C_1,\ldots,C_{k-1}\rangle:\ini(C_k)^{\infty}$ in Line \ref{line:StrongTriDec-Gsat} instead of the ideal quotient of $\langle C_1,\ldots,C_{k-1}\rangle$ by $\ini(C_k)$ (see \cite[Algorithm 1-Line 19]{DBLP:conf/casc/DongM19}).
And Algorithm \ref{alg:StrongTriDec} can detect  whether the input system is zero-dimensional.
\end{remark}

\vspace{-1.5mm}
\vspace{-2.5mm}
\section{SFSTD}\label{sec:strongsftridec}
\vspace{-0.5mm}

A popular method for computing square-free/regular chains is the method of relatively simplicial decomposition (see \cite[Chapter 2]{xia2016automated} for more details), which is based on subresultant computation and pseudo-division.
In this section, we discuss how to compute SFSTD by means of Gr\"obner bases.
We propose Algorithm \ref{alg:SubStrongSfTriDec} for SFSTD of strong chains in Section \ref{subsec:StrongSfTriDecOfFullPureChain} and Algorithm \ref{alg:StrongSfTriDec} for SFSTD of general zero-dimensional systems in Section \ref{subsec:StrongSfTriDecOfZeroDim}.  

\vspace{-1.5mm}

\begin{definition}\label{def:sfstd}
Let $F\subseteq\Q[\vX]$ be a zero-dimensional system. 
A \emph{square-free strong  triangular decomposition} (SFSTD) of $F$ is an STD $\{\mT_1,\ldots,\mT_s\}$ of $F$, where $\mT_i$ is square-free for $i=1,\ldots,s$.
\end{definition}

\vspace{-1.5mm}
\vspace{-2.5mm}
\subsection{Computing SFSTD of Strong Chains}\label{subsec:StrongSfTriDecOfFullPureChain}
\vspace{-0.5mm}
Algorithm \ref{alg:SubStrongSfTriDec} computes an SFSTD of any strong chain $\mT\subseteq\Q[x_1,\ldots,\allowbreak x_n]$.
Let $\Phi$ be a set of strong chains for SFSTD (initialized as $\{\mT\}$), and $ans$ be a set of computed square-free strong chains (initialized as $\emptyset$).
Every loop step, we pick a strong chain $\mP=[P_1,\ldots,P_n]$ from $\Phi$ and remove it from $\Phi$, until $\Phi$ is empty.
\begin{enumerate}[1.]
    \item If $\mP$ is square-free, then we add $\mP$ to $ans$. 
    \item If $\mP$ is not square-free, then we compute $k~(k\ge1)$ which is the smallest integer such that $[P_1,\ldots,P_k]$ is not square-free. 
    If $k=1$, then $\Phi$ is updated with the strong chains $[\xi_1,P_2,\ldots,P_n],\ldots,\allowbreak[\xi_m,P_2,\ldots,P_n]$, where $\xi_i$ is an irreducible factor of $P_1$.
    If $k>1$, we compute the reduced Gr\"obner basis $G_{sat}$ of $\langle P_1,\ldots,P_{k}\rangle:\sep(P_k)^{\infty}$ w.r.t. any monomial ordering.
    And, by Algorithm \ref{alg:StrongTriDec}, we compute an STD of $\mP\cup\{\sep(P_k)\}$ and an STD of $\mP\cup G_{sat}$.
    $\Phi$ is updated with the strong chains in the two STD.
\end{enumerate}


\begin{algorithm}[t]
\scriptsize
\DontPrintSemicolon
\LinesNumbered
\SetKwInOut{Input}{Input}
\SetKwInOut{Output}{Output}
\Input{a strong chain $\mT=[T_1,\ldots,T_n]\subseteq\Q[\vX]$ and the vector $\vX$}
\Output{$ans=\{\mQ_1,\ldots,\mQ_s\}$, a finite set of square-free strong chains such that
\[  \V(\mT)~=~\bigcup_{\mQ_i\in ans}\V(\mQ_i)~{\rm and}~\V(\mQ_i)\cap\V(\mQ_j)=\emptyset~{\rm for~any~}i\ne j
\]
}
\caption{{\bf SubSFSTD} (Sub-Algorithm of Algorithm \ref{alg:StrongSfTriDec})}\label{alg:SubStrongSfTriDec}
\BlankLine
$ans\leftarrow\emptyset,~\Phi\leftarrow\{\mT\}$\;
\While{$\Phi\ne\emptyset$}
{
 Choose $\mP=[P_1,\ldots,P_n]$ from $\Phi$ and set $\Phi\leftarrow\Phi\setminus\{\mP\}$\label{algline:subssftd:choose}\; 
 \eIf{$\mP$ is square-free\label{line:StrongSfTriDec-sf}}
 {$ans\leftarrow ans\cup\{\mP\}$\label{line:StrongSfTriDec-add}}
 {$P_k\leftarrow$ the first polynomial that makes $[P_1,\ldots,P_k]$ not  square-free\;
  \eIf{$k=1$}
  {
   $\xi_1,\ldots,\xi_m\leftarrow$ all irreducible factors of $P_1$\;
   \# Here, $\xi_i\in\Q[\vX]$ for $i=1,\ldots,m$.\;
   $\Phi\leftarrow\Phi\cup\{[\xi_1,P_2,\ldots,P_n],\ldots,[\xi_m,P_2,\ldots,P_n]\}$\label{line:StrongSfTriDec-Factor}\;
  }
  {
   $G_{sat}\leftarrow$ the reduced Gr\"obner basis of $\langle P_1,\ldots,P_k\rangle:\sep(P_k)^{\infty}$ w.r.t. any monomial ordering\label{line:StrongSfTriDec-GB}\; 
   $\Phi\leftarrow\Phi\cup{\bf STD}(\mP\cup\{\sep(P_k)\},\vX)
   \cup{\bf STD}(\mP\cup G_{sat},\vX)$\label{line:StrongSfTriDec-Spilt}
  }
 }
}
\Return{$ans$}
\end{algorithm}

Algorithm \ref{alg:SubStrongSfTriDec} is illustrated on the following example.
\vspace{-1.5mm}

\begin{example}
Consider the strong chain $\mT=[T_1,T_2]=[x^2-1, y^2-2xy+1]\subseteq\Q[x,y]$ with $x<y$.
$\mT$ is not square-free and $T_2$ is the first polynomial that makes it not square-free.
Since $\langle T_1\rangle:\sep(P_2)^\infty=\langle1\rangle$ where $\sep(P_2)=2y-2x$, we only compute an STD of $\mP\cup\{\sep(P_k)\}$ by Algorithm \ref{alg:StrongTriDec}, which is $\{[x^2-1,y-x]\}$.
$\Phi$ is updated with the strong chain $[x^2-1,y-x]$.
Because the strong chain is square-free, 
Algorithm \ref{alg:StrongSfTriDec} terminates with $\Phi=\emptyset$ and $\{[x^2-1,y-x]\}$ is an SFSTD of $\mT$.
\end{example}
\vspace{-1.5mm}

\vspace{-1.5mm}
\begin{theorem}\label{thm:SubStrongSfTriDecIsCorrect}
Algorithm \ref{alg:SubStrongSfTriDec} terminates correctly.
\end{theorem}
\vspace{-3mm}
\begin{proof}
(Correctness) 
We choose a strong chain $\mP$ from $\Phi$ in every loop.
If $\mP$ is square-free, then it is added to the output set in Line \ref{line:StrongSfTriDec-add}.
Otherwise, $\Phi$ is updated with some strong chains in Line \ref{line:StrongSfTriDec-Factor} or Line \ref{line:StrongSfTriDec-Spilt}
(note that since $\mP$ is zero-dimensional by Proposition \ref{prop:PureChainIsGro} (\ref{item:PureChainIsGro4}), Algorithm \ref{alg:StrongTriDec} does not return FAIL in Line \ref{line:StrongSfTriDec-Spilt}).
So, it is obvious that we only need to prove that $\V(\mP)=\V(\mP\cup\{\sep(P_k)\})\cup\V(\mP\cup G_{sat})$ and $\V(\mP\cup\{\sep(P_k)\})\cap\V(\mP\cup G_{sat})=\emptyset$, where $P_k~(k>1)$ is the first polynomial that makes $[P_1,\ldots,P_k]$ not square-free and $G_{sat}$ is a Gr\"obner basis of $\langle P_1,\ldots,P_k\rangle:(\sep(P_k))^{\infty}$.
Note that $\{P_1,\ldots,P_k\}$ is also zero-dimensional by Proposition \ref{prop:PureChainIsGro} (\ref{item:PureChainIsGro4}).
Then, similar to the correctness proof of Theorem \ref{thm:algStrongTriDecIsCorrect}, we have $\V(\mP\cup G_{sat})=\V(\mP)\setminus\V(\sep(P_k))$ and $\V(\mP\cup\{\sep(P_k)\})=\V(\mP)\cap\V(\sep(P_k))$.

(Termination) 
The termination is equivalent to that every ideal generated by the added strong chain in Line \ref{line:StrongSfTriDec-Factor} or Line \ref{line:StrongSfTriDec-Spilt} is strictly larger than that generated by the removed one.
Note that it is clear for every strong chain added in Line \ref{line:StrongSfTriDec-Factor}.
Thus, by Corollary \ref{coro:STD}, we only need to prove that $\langle\mP\cup\{\sep(P_k)\}\rangle$ and $\langle\mP\cup G_{sat}\rangle$ are both strictly larger than $\langle\mP\rangle$.

Firstly, we prove  $\langle\mP\rangle\subsetneq\langle\mP\cup\{\sep(P_k)\}\rangle$.
Since $\mP=[P_1,\ldots,P_n]$ is a strong chain, by Proposition \ref{prop:PureChainIsGro}, 
$\mP$ is a LEX Gr\"obner basis, and $\lt(P_i)=c_ix_i^{j_i}$ where $c_i\in\Q\setminus\{0\}$ and $j_i\ge1$. 
Then, $\lt(\langle\mP\rangle)=\langle\lt(\mP)\rangle=\langle c_1x_1^{j_1},\ldots,c_nx_n^{j_n}\rangle$, and either $\lt(\sep(P_k))=j_kc_kx_k^{j_{k}-1}$ ($j_k>1$) or $\lt(\sep(P_k))=c_k$ ($j_k=1$). 
So, $\lt(\sep(P_k))\not\in\lt(\langle\mP\rangle)$. 
Thus, $\sep(P_k)\notin\langle\mP\rangle$.
The proof is completed.


Secondly, we prove  $\langle\mP\rangle\subsetneq\langle\mP\cup G_{sat}\rangle$, which is equivalent to proving
that there exists 
$g\in\langle G_{sat}\rangle\setminus\langle\mP\rangle$.
Note that $[P_1,\ldots,P_{k-1}]$ is square-free, but $[P_1,\ldots,P_{k}]$ is not. 
So,  $\res(\sep(P_k),[P_1,\ldots,P_{k}])=0$.
Then, because $[P_1,\ldots,P_k]$ is a strong chain in $\Q[x_1,\ldots,x_k]$, 
by Proposition \ref{prop:StrongChainOr},
there exists $g\in\CC[x_1,\ldots,x_{k}]\setminus\langle P_1,\ldots,P_k\rangle$ such that $g\cdot\sep(P_k)\in\langle P_1,\ldots,P_k\rangle$. 
Recall that $\langle G_{sat}\rangle=\langle P_1,\ldots,P_k\rangle:(\sep(P_k))^{\infty}$.
So, $g\in\langle G_{sat}\rangle$.
It remains to prove $g\notin\langle\mP\rangle$.
Because $\mP$ is a LEX Gr\"obner basis, we have $\langle\mP\rangle\cap\CC[x_1,\ldots,x_{k}]=\langle P_1,\ldots,P_k\rangle$.
Note that $g\in\CC[x_1,\ldots,x_{k}]\setminus\langle P_1,\ldots,P_k\rangle$.
So, $g\not\in\langle\mP\rangle$. 
\end{proof}

\vspace{-1.5mm}
\vspace{-2.5mm}
\subsection{Computing SFSTD}\label{subsec:StrongSfTriDecOfZeroDim}
\vspace{-0.5mm}

For any zero-dimensional system $F\subseteq\Q[\vX]$, Algorithm \ref{alg:StrongSfTriDec} computes an SFSTD of $F$.
The process of the computation is as follows.
We first compute an STD of $F$ by Algorithm \ref{alg:StrongTriDec} and then compute an SFSTD of every strong chain in the STD by Algorithm \ref{alg:SubStrongSfTriDec}. 
The union of all SFSTD is an SFSTD of $F$.
If the input system is not zero-dimensional, Algorithm \ref{alg:StrongSfTriDec} returns FAIL in Line \ref{line:algSFSTD-FAIL}.

\begin{algorithm}[t]
\scriptsize
\DontPrintSemicolon
\LinesNumbered
\SetKwInOut{Input}{Input}
\SetKwInOut{Output}{Output}
\Input{a nonempty finite set $F\subseteq\Q[\vX]\setminus\{0\}$ and the vector $\vX$}
\Output{$ans=\{\mT_1,\ldots,\mT_s\}$, a finite set of square-free strong chains such that
\[  \V(F)~=~\bigcup_{\mT_i\in ans}\V(\mT_i)~{\rm and}~\V(\mT_i)\cap\V(\mT_j)=\emptyset~{\rm for~any~}i\ne j,
\]
if $F$ is zero-dimensional;
FAIL otherwise
}
\caption{{\bf SFSTD}}\label{alg:StrongSfTriDec}
\BlankLine
$\Lambda\leftarrow{\bf STD}(F,\vX)$\label{line:SFSTD-STD}\;
\eIf{$\Lambda$ = FAIL}
{\Return{FAIL}\label{line:algSFSTD-FAIL}}
{
\Return{$\bigcup_{\mP\in\Lambda}{\bf SubSFSTD}(\mP,\vX)$\label{algline:SFSTD:cup}} 
}
\end{algorithm}
\vspace{-1.5mm}
\begin{theorem}\label{thm:algStrongSfTriDecIsCorrect}
Algorithm \ref{alg:StrongSfTriDec} terminates correctly.
\end{theorem}
\vspace{-3mm}
\begin{proof}
It is obvious by Theorem \ref{thm:algStrongTriDecIsCorrect} and Theorem \ref{thm:SubStrongSfTriDecIsCorrect}.
\end{proof}

\vspace{-1.5mm}
\vspace{-2.5mm}
\section{Arithmetic Complexity Analysis}\label{sec:complexity}
\vspace{-0.5mm}

In the section, we analyze the complexity of our algorithms.
Here, we only consider arithmetic complexity which counts the number of field operations (not bit operations). 
We first introduce the concept of multiplicity and some results we will use later.

For any point $p=(a_1,\ldots,a_n)\in\CC^n$,
we denote by
$\CC[\vX]_{p}$ the set $\left\{\frac{f}{g}\mid\allowbreak f,g\allowbreak\in\CC[\vX],\allowbreak g(a_1,\ldots,a_n)\neq 0 \right\}.$
For any ideal $I\subseteq\CC[\vX]$, we denote by $\dim(I)$ the dimension of the $\CC$-vector space $\CC[\vX]/I$.
For any $P\subseteq\Q[\vX]$, define $\dim(P):=\dim(\langle P\rangle)$ and denote by $\deg(P)$ the maximum degree of elements of $P$.  
\vspace{-1.5mm}
\begin{definition}[{\cite[Chap. 4. Def. 2.1]{Cox_Little_OShea_2005}}]\label{def:RootMulti}
Let $I\subseteq\CC[\vX]$ be a zero-dimensional ideal and $p\in\V(I)$.
The \emph{multiplicity} of $p$, denoted $\m{I}(p)$, is the dimension of the $\CC$-vector space $\CC[\vX]_{p}/I\CC[\vX]_{p}$.
\end{definition}
\vspace{-1.5mm}

\vspace{-1.5mm}

\begin{theorem}[{\cite[Chap. 4. Cor. 2.5]{Cox_Little_OShea_2005}}]\label{thm:dim-multi}
Let $I\subseteq\CC[\vX]$ be a zero-dimensional ideal.
We have $\dim(I)=\sum_{\alpha\in\V(I)}\m{I}(\alpha)$.
\end{theorem}

\vspace{-3mm}

\begin{theorem}
[{\cite[Prop. 8.1]{hashemi2005complexity}}, {\cite[Thm. 3]{lakshman1991single}}]
\label{thm:GBCOM}
Let $P\subseteq\Q[x_1,\ldots,x_n]$ be a zero-dimensional system,
and $G$ be the reduced Gr\"obner basis of $\langle P\rangle$ w.r.t. any monomial ordering.
Then, 
\begin{enumerate}[(a)]
    \item\label{item:thm:GBCOM1} $\deg(G)\le\dim(P)$, and
    \item\label{item:thm:GBCOM2} the arithmetic complexity of computing $G$, denoted $\cgblex(n,\deg(P))$, can be polynomial in $\deg(P)^n$.
\end{enumerate}
\end{theorem}
\vspace{-1.5mm}
Remark that in Theorem \ref{thm:GBCOM}, $\cgblex(n, \deg(P))$ depends on algorithms used for computing Gr\"obner bases.


\vspace{-2.5mm}
\subsection{Complexity of Algorithm \ref{alg:StrongTriDec}}\label{subsec:STDcomplexity}
\vspace{-0.5mm}

Let $F\subseteq\Q[x_1,\ldots,x_n]$ be a zero-dimensional system, $d:=\deg(F)$
and $D:=\max(d,\dim(F))$ in Section \ref{subsec:STDcomplexity} and Section \ref{subsec:SFSTDcomplexity}.  
For convenience, we assume $\V(F)\ne\emptyset$.

\vspace{-1.5mm}
\begin{remark}\label{remark:Bezout}
    By B\'ezout's theorem, $D\leq d^n$.
\end{remark}
\vspace{-3mm}

\begin{theorem} \label{thm:complexityofSTD}
Algorithm \ref{alg:StrongTriDec} computes an STD of $F$ with the arithmetic complexity
\begin{align}\label{eq:STDcomplexity}
    (2\dim(F)-1)\cdot(\cgblex(n,D)+\cgblex(n,\dim(F))),
\end{align}
which can be polynomial in $D^n$. 
\end{theorem}
\vspace{-1.5mm}

In order to prove Theorem \ref{thm:complexityofSTD}, we prepare some lemmas.

Given the input $F$, suppose that Algorithm \ref{alg:StrongTriDec} terminates with $N~(N\ge1)$ times of loops.
Let $P_i$ be the picked polynomial set in the $i$-th loop. 
Consider a binary tree $T$ with $P_1,\ldots,P_{N}$ as nodes.
If $\langle P_i\rangle\neq \langle1\rangle$  and the  W-characteristic set of $P_i$ is not strong, the node $P_i$ has two child nodes $P_{i_1}=\{G\cup\{\ini(C_k)\}\}$ and $P_{i_2}=\{G\cup G_{sat}\}$ (see Line \ref{line:StrongTriDec-Spilt}), where 
$i<i_j\le N$
for $j=1,2$. 
Otherwise, the node $P_i$ is a leaf.
We denote by $\child(P_i)$ the set of all child nodes of $P_i$ (if $P_i$ is a leaf, $\child(P_i)=\emptyset$).
The root node of tree $T$ is $P_1=F$.

For each node $P_i$ of tree $T$, we define a value:
\begin{align}\label{eq:defvalue}
\Value(P_i)=
\left\{
\begin{array}{cc}
\dim(P_i), & \V(P_i)\neq\emptyset,\\
1,     &  \V(P_i)=\emptyset.
\end{array}
\right.
\end{align}
Note that $\V(P_i)=\emptyset$ if and only if $\dim(P_i)=0$, and thus $\dim(P_i)\le\Value(P_i)$.

\vspace{-1.5mm}

\begin{lemma}\label{lemma:SubsetIdeal}
If $\langle F\rangle\subsetneq\langle G\rangle\subseteq\CC[\vX]$, then 
\begin{enumerate}[(a)]
    \item \label{item:lemma:SubsetIdeal1}$\m{\langle G \rangle}(p)\le \m{\langle F \rangle}(p)$ for any $p\in\V(G)$, and
    \item \label{item:lemma:SubsetIdeal2}$\dim(G)<\dim(F)$.
\end{enumerate}
\end{lemma}
\vspace{-1.5mm}
\vspace{-1.5mm}

\begin{proof}
It is obvious by the definitions of multiplicity and dimension.
\end{proof}
\vspace{-1.5mm}
\vspace{-1.5mm}

\begin{lemma} \label{lem:STDTREE}
For the tree $T$, we have
\begin{align}
\sum_{P_{i_j}\in\child(P_i)}\Value(P_{i_j})&\leq \dim(P_i),\label{eq:lem-STDTREE1}\\
\sum_{P_i \text{is a leaf of } T}\Value(P_i)&\leq \dim(F).\label{eq:lem-STDTREE2}
\end{align}
\end{lemma}
\vspace{-1.5mm}
\vspace{-1.5mm}
\begin{proof}
We prove \eqref{eq:lem-STDTREE1} first.
If $P_i$ is a leaf, it is clear.
Otherwise, we have $\langle P_i\rangle\ne\langle1\rangle$, i.e., $\V(P_i)\ne\emptyset$.
Note that $\V(P_i)=\V(P_{i_1})\cup\V(P_{i_2})$ by the proof of Theorem \ref{thm:algStrongTriDecIsCorrect}.
Then, at least one variety of child nodes is not $\emptyset$.
We also note that $\langle P_i\rangle\subsetneq\langle P_{i_j} \rangle$ for $j=1,2$ and $V(P_{i_1})\cap V(P_{i_2})=\emptyset$ by the proof of Theorem \ref{thm:algStrongTriDecIsCorrect}.
Thus, if neither $\V(P_{i_1})$ nor $\V(P_{i_2})$ is an empty set, then by Theorem \ref{thm:dim-multi} and Lemma \ref{lemma:SubsetIdeal} (\ref{item:lemma:SubsetIdeal1}),
\begin{align}
    \sum_{P_{i_j}\in\child(P_i)}\Value(P_{i_j})&=\sum_{j=1,2}\dim(P_{i_j})\nonumber=\sum_{j=1,2}\sum_{p\in\V(P_{i_j})}\m{\langle P_{i_j} \rangle}(p)\nonumber\\
    &\le\sum_{p\in\V(P_{i})} \m{\langle P_{i} \rangle}(p)=\dim(P_i).\label{eq:treeconclu1}
\end{align}
If $\V(P_{i_1})=\emptyset$ and  $\V(P_{i_2})\ne\emptyset$, then by Lemma \ref{lemma:SubsetIdeal} (\ref{item:lemma:SubsetIdeal2}),
\begin{align}\label{eq:treeconclu2}
\sum_{P_{i_j}\in\child(P_i)}\Value(P_{i_j})=1+\dim(P_{i_2})\le\dim(P_{i}).
\end{align}
By \eqref{eq:treeconclu1} and \eqref{eq:treeconclu2}, \eqref{eq:lem-STDTREE1} is proved.
Then, it is clear that \eqref{eq:lem-STDTREE2} holds by induction.
\end{proof}
\vspace{-1.5mm}
\vspace{-1.5mm}

\begin{lemma}\label{lemma:treedeg}
For every node $P_i~(i=1,\ldots,N)$ of tree $T$, let $G_i$ be the reduced LEX Gr\"obner basis of $\langle P_i\rangle$. Then,
\begin{enumerate}[(a)]
    \item\label{item:lemma:treedeg1} $\deg(G_i)\le\dim(F)$,
    \item\label{item:lemma:treedeg2} $\deg(P_i)\le\max(d,\dim(F))$.
\end{enumerate}
\end{lemma}
\vspace{-1.5mm}
\vspace{-1.5mm}

\begin{proof}
(\ref{item:lemma:treedeg1}) By Theorem \ref{thm:GBCOM}, $\deg(G_i)\allowbreak\le\dim(P_i)$. 
Then, we prove $\dim(P_i)\le\dim(F)$ by induction on $i$.
It is clear for $i=1$. 
Assume it holds for $1,\ldots,i-1$. 
By Lemma \ref{lem:STDTREE}, $\Value(P_i)\le \dim(P_{i^*})$, where $P_{i^*}$ is the parent of $P_i$.
Since $i^*<i$, by the assumption, $\dim(P_{i^*})\le\dim(F)$. 
Thus, $\dim(P_i)\le\Value(P_i) \le\dim(P_{i^*})\le\dim(F)$.

(\ref{item:lemma:treedeg2}) The maximum degree of elements of $P_1=F$ is $d$.
Note that each node $P_j~(2\le j\le N)$ has a parent $P_{j^*}~(1\le j^*< j)$.
So, $P_j$ is one of $G\cup\{\ini(C_k)\}$ and $G\cup G_{sat}$ in Line \ref{line:StrongTriDec-Spilt}, where $G$ is the reduced LEX Gr\"obner basis of $\langle P_{j^*}\rangle$.
We only need to prove that $\deg(G\cup\{\ini(C_k)\})\le\dim(F)$ and $\deg(G\cup G_{sat})\le\dim(F)$.
By (\ref{item:lemma:treedeg1}), we have $\deg(G)\le\dim(F)$.
Then, since $C_k\in G$, 
the degree of $\ini(C_k)$ is at most $\dim(F)$.
It remains to prove $\deg(G_{sat})\le\dim(F)$.
By Theorem \ref{thm:GBCOM} (\ref{item:thm:GBCOM1}), $\deg(G_{sat})\le\dim(G_{sat})$.
Because $\langle C_1,\ldots,C_{k-1}\rangle\subseteq\langle G_{sat}\rangle$, by Lemma \ref{lemma:SubsetIdeal} (\ref{item:lemma:SubsetIdeal2}), $\dim(G_{sat})\le\dim(C_1,\ldots,C_{k-1})$.
By Lemma \ref{lemma:WcharIsStrong},
$\langle C_1,\ldots,C_{k-1}\rangle=\langle G\cap\CC[x_1,\ldots,x_{k-1}]\rangle$.
So, $\dim(\{C_1,\ldots,\allowbreak C_{k-1}\})$ is equal to the dimension of $\CC[x_1,\ldots,x_{k-1}]/(\langle G\rangle\cap\CC[x_1,\ldots,\allowbreak x_{k-1}])$.
Note that 
$\CC[x_1,\allowbreak\ldots,x_{k-1}]/\allowbreak(\langle G\rangle\cap\CC[x_1,\ldots,x_{k-1}])$ is equal to the quotient ring $\CC[x_1,\ldots,x_n]/\langle G\rangle$ limited on $\CC[x_1,\ldots,x_{k-1}]$.
Then, the dimension 
is at most $\dim(G)$, i.e., $\dim(P_{j^*})$.
Note that $\dim(P_{j^*})\le\dim(F)$ by the proof of (\ref{item:lemma:treedeg1}).
We complete the proof.
\end{proof}
\vspace{-1.5mm}
\vspace{-1.5mm}

\begin{lemma}\label{lemma:STDN}
    The number of leaves of tree $T$ is at most $\dim(F)$.
    The number of nodes of tree $T$ is at most $2\dim(F)-1$, i.e., $N\leq 2\dim(F)-1$.
\end{lemma}
\vspace{-1.5mm}
\vspace{-1.5mm}

\begin{proof}
Note that $\Value(P_i)\ge1$ for $i=1,\ldots,N$.
Then, by \eqref{eq:lem-STDTREE2} of Lemma \ref{lem:STDTREE}, the number of leaves is at most $\dim(F)$.
Note that every node is either a leaf or has two child nodes.
So, the number of nodes is at most $2\dim(F)-1$.
\end{proof}
\vspace{-1.5mm}

\textbf{Proof of Theorem \ref{thm:complexityofSTD}.}
By Lemma \ref{lemma:STDN}, the number of times of loops $N\le2\dim(F)\allowbreak-1$. 
In each loop, the most complicated computation is Gr\"obner bases computation in Line \ref{line:StrongTriDec-ReGB} and Line \ref{line:StrongTriDec-Gsat}.
By Lemma \ref{lemma:treedeg} (\ref{item:lemma:treedeg2}), the computation in Line \ref{line:StrongTriDec-ReGB} has the complexity $\cgblex(n,D)$.
Note that the saturated ideal in Line \ref{line:StrongTriDec-Gsat} is equal to $\langle C_1,\ldots,C_{k-1},1-\tau\ini(C_k)\rangle\cap\CC[x_1,\ldots,x_{k-1}]$, where $\tau$ is a new variable.
So, by Lemma \ref{lemma:treedeg} (\ref{item:lemma:treedeg1}), the complexity of the computation in Line \ref{line:StrongTriDec-Gsat} is
$\cgblex(k,\dim(F))$.  Note that $k\leq n$.
Thus, \eqref{eq:STDcomplexity} is proved.
By Theorem \ref{thm:GBCOM} (\ref{item:thm:GBCOM2}), \eqref{eq:STDcomplexity} can be polynomial in $D^n$.   \hfill$\square$\par

\vspace{-2.5mm}
\subsection{Complexity of Algorithms \ref{alg:SubStrongSfTriDec} \& \ref{alg:StrongSfTriDec}}\label{subsec:SFSTDcomplexity}
\vspace{-0.5mm}

\begin{theorem}\label{thm:SFSTDcomplety}
An SFSTD of a reduced strong chain $\mT\subseteq\Q[\vX]$ can be computed by Algorithm \ref{alg:SubStrongSfTriDec} within a complexity of polynomial in $\dim(\mT)^n$.
An SFSTD of $F\subseteq\Q[\vX]$ can be computed by Algorithm \ref{alg:StrongSfTriDec} within a complexity of polynomial in $d^{n^2}$.
\end{theorem}

\vspace{-1.5mm}

To prove Theorem \ref{thm:SFSTDcomplety}, we prepare some lemmas first.

Given a reduced strong chain $\mT$, suppose that Algorithm \ref{alg:SubStrongSfTriDec} terminates with $M~(M\ge1)$ times of loops.
Let $\mP_i$ be the picked reduced strong chain in the $i$-th loop (see Corollary \ref{coro:STD}).
Consider a tree $\tilde{T}$ with some nodes $\mP_1,\ldots,\mP_{M}$ and some other nodes $\{1\}$.
If $\mP_i$ is not square-free, 
then the node $\mP_i$ has one or more reduced strong chains in Line \ref{line:StrongSfTriDec-Factor} or Line \ref{line:StrongSfTriDec-Spilt} as child nodes.
Otherwise,  the node $\mP_i$ has no child node. 
If $\mP_i$ has and only has one child node, let the set $\{1\}$ be its second child node.
The root node of tree $\tilde{T}$ is $\mP_1=\mT$.
We also define a value of every node $\mP_i$ as in \eqref{eq:defvalue}.
\vspace{-1.5mm}
\begin{lemma}\label{lemma:TildeTreeNodeDeg}
For every node $\mP$ of tree ${\tilde T}$, $\deg(\mP)\le\dim(\mP)$.   
\end{lemma}
\vspace{-3mm}
\begin{proof}
If $\mP=\{1\}$, the conclusion is clear.
Otherwise, $\mP$ is a reduced strong chain.
Then, by Proposition \ref{prop:PureChainIsGro} and Theorem \ref{thm:GBCOM} (\ref{item:thm:GBCOM1}), we complete the proof.
\end{proof}
\vspace{-3mm}
\begin{lemma}  \label{lem:sfstd-tree}
For the tree ${\tilde T}$, we have $\sum\limits_{\mP_{i_j}\in\child(\mP_i)}\Value(\mP_{i_j})\leq \dim(\mP_i)$ and  $\sum\limits_{\mP_i \text{is a leaf of } {\tilde T}}\Value(\mP_i)\leq\dim(\mT)$.
\end{lemma}
\vspace{-3mm}
\begin{proof}
Note that in Line \ref{line:StrongSfTriDec-Factor},
$\langle P_1,\ldots,P_n\rangle\subsetneq\langle\xi_i,P_2,\ldots,P_n\rangle$ for $i=1,\ldots,m$, $\V(P_1,\allowbreak\ldots,P_n)=\cup_{i=1}^{m}\V(\xi_i,P_2,\ldots,P_n)$ and $\V(\xi_i,P_2,\ldots,\allowbreak P_n)\cap\V(\xi_j,P_2,\ldots,\allowbreak P_n)=\emptyset$ for $i\ne j$.
Then, it is similar to the proof of Lemma \ref{lem:STDTREE}.
\end{proof}
\vspace{-3mm}
\begin{lemma}\label{lemma:SSTDN}
The number of nodes of tree $\tilde{T}$ is at most $2\dim(\mT)-1$ which implies $M\leq 2\dim(\mT)-1$.
\end{lemma}
\vspace{-3mm}
\begin{proof}
Note that if a node is not a leaf, then it has at least two child nodes.
Then, it is similar to the proof of Lemma \ref{lemma:STDN}.
\end{proof}
\vspace{-2mm}
\textbf{Proof of Theorem \ref{thm:SFSTDcomplety}.}
Firstly, we analyze the complexity of Algorithm \ref{alg:SubStrongSfTriDec}.
By Lemma \ref{lemma:SSTDN}, the number of loop steps
$M\le2\dim(\mT)-1$.
In each loop, the most complicated computation is in Line \ref{line:StrongSfTriDec-GB} and Line \ref{line:StrongSfTriDec-Spilt}.
Note that in Line \ref{line:StrongSfTriDec-GB}, $\langle G_{sat}\rangle=\langle P_1,\ldots,P_k,1-\tau\sep(P_k)\rangle\cap\CC[x_1,\ldots,x_k]$, where $k\le n$, $\tau$ is a new variable and $G_{sat}$ is a reduced Gr\"obner basis.
So, by Lemma \ref{lemma:TildeTreeNodeDeg}, the complexity of computing $G_{sat}$ is $\cgblex(n+1,\dim(\mP))$, where $\mP$ is the reduced strong chain chose in Line \ref{algline:subssftd:choose}.
By Theorem \ref{thm:GBCOM} (\ref{item:thm:GBCOM2}), it can be polynomial in $\dim(\mP)^n$.
It remains to analyze the complexity of computation in Line \ref{line:StrongSfTriDec-Spilt}.
Similar to the proof of Lemma \ref{lemma:treedeg}, $\dim(\mP)\le \dim(\mT)$ and $\deg(G_{sat})\leq \dim(\mP)$.
And by Lemma \ref{lemma:TildeTreeNodeDeg},
$\deg(\mP)\le\dim(\mP)$. 
Thus, it is clear that $\deg(\mP\cup\{\sep(P_k)\})$, $\dim(\mP\cup\{\sep(P_k)\})$, $\deg(\mP\cup G_{sat})$ and $\dim(\mP\cup G_{sat})$ are all less than or equal to $\dim(\mT)$.
Then, by Theorem \ref{thm:complexityofSTD}, the complexity 
can be polynomial in $\dim(\mT)^n$.
Therefore, after multiplying $M$, the complexity of Algorithm \ref{alg:StrongSfTriDec} can still be polynomial in $\dim(\mT)^{n}$.

Secondly, we analyze the complexity of Algorithm \ref{alg:StrongSfTriDec}.
By Theorem \ref{thm:complexityofSTD}, the complexity of the calculation in Line \ref{line:SFSTD-STD} can be polynomial in $D^n$.
Suppose the reduced strong chains (see Corollary \ref{coro:STD}) computed in Line \ref{line:SFSTD-STD} are $\mT_1,\ldots,\mT_t$.
By Lemma \ref{lemma:STDN}, $t\le\dim(F)$.
By \eqref{eq:lem-STDTREE2} of Lemma \ref{lem:STDTREE}, $\dim(\mT_i)\le\Value(\mT_i)\le\dim(F)$.
Thus, by the conclusion in the above paragraph, the complexity of the calculation in Line \ref{algline:SFSTD:cup} can be polynomial in $\dim(F)^n$.
Then, the complexity of Algorithm \ref{alg:StrongSfTriDec} can be polynomial in $\max(D^n,\dim(F)^n)=D^{n}$. 
Since $D<d^n$ (see Remark \ref{remark:Bezout}), the complexity can be polynomial in $d^{n^2}$. \hfill$\square$\par

Recall that Theorem \ref{thm:complexityofSTD} and Theorem \ref{thm:SFSTDcomplety} talk about arithmetic complexity without analyzing the growth of the size of coefficients.
In fact, the size of the coefficients in the algorithms may increase very fast.

\vspace{-3mm}
\section{Two Applications of SFSTD}\label{sec:applications}
\vspace{-1mm}

In the section, we present two applications of SFSTD:
real solution isolation and computing radicals.
Given a zero-dimensional system $F\subseteq\Q[\vX]$,
we first compute an SFSTD $\{\mT_1,\ldots,\mT_s\}$ of 
$F$ by Algorithm \ref{alg:StrongSfTriDec}.

In order to compute the isolating cubes of real solutions of $F$, we compute the isolating cubes of every square-free strong chain $\mT_i$ by \cite[Algorithm {NREALZERO}]{xia2006real}. 
The method is called NRSI in Section \ref{sec:experiments}.

We denote by $\sqrt{I}$ the radical of an ideal $I\subseteq\CC[\vX]$.
We claim that $\sqrt{\langle F\rangle}=\bigcap_{i=1}^{s}\langle\mT_i\rangle$.
The proof of the claim is as follows.
Since $\{\mT_1,\ldots,\mT_s\}$ is an SFSTD of $F$, we have
$\V(F)=\bigcup_{i=1}^{s}\V(\mT_i)$.
Then, $\V(F)=\V(\bigcap_{i=1}^{s}\langle\mT_i\rangle)$.
So, $\sqrt{\langle F\rangle}=\bigcap_{i=1}^{s}\sqrt{\langle\mT_i\rangle}$.
Note that every $\mT_i$ is a square-free strong chain.
Then, by \cite[Corollary 3.3]{boulier2006well} and by Proposition \ref{prop:PureChainIsGro} (\ref{item:PureChainIsGro1}), $\langle\mT_i\rangle$ is radical.
Thus, we complete the proof.
The method for computing $\sqrt{\langle F\rangle}$ by the intersection of ideals is called IRA in Section \ref{sec:experiments}.

\vspace{-2mm}
\section{Experiments}\label{sec:experiments}
\vspace{-0.5mm}

We implemented Algorithm \ref{alg:StrongSfTriDec}, the methods NRSI and IRA with {\tt Maple2021}, where we use the {\tt Maple} command {\tt Groebner[Basis]} for computing Gr\"obner bases in Algorithm \ref{alg:StrongTriDec}--Line \ref{line:StrongTriDec-ReGB}\&Line \ref{line:StrongTriDec-Gsat} and Algorithm \ref{alg:SubStrongSfTriDec}--Line \ref{line:StrongSfTriDec-GB}.

In the section, we explain implementation details and show the experimental results of partial testing examples.
All testing examples, code and experimental results are available online via:{ \bf \url{https://github.com/lihaokun/StrongSfTriDec}}.
All tests were conducted on 16-Core Intel Core i7-12900KF@3.20GHz
with 128GB of memory and Windows 11.

\vspace{-3mm}
\subsection{Description of the Experimentation}\label{subsec:description}
\vspace{-0.5mm}

Testing examples are collected from the literatures \cite{boulier2014real, wang1996solving, xia2006real} and the website \url{http://homepages.math.uic.edu/~jan/demo.html}.
We just get rid of the ones that are repeated or not zero-dimensional.
Owing to space constraints, we only present $44$ ``difficult'' examples (total $151$ examples) in Table \ref{tab:timing}.
Timings are in seconds.
``OT'' means out of the timing 3600 seconds, and ``LOSS'' means kernel connection lost during calculation of {\tt Maple}.
The column ``sys'' denotes the name of the polynomial system.
The column ``$n/d$'' stands for the number of variables/the maximum degree of elements in the system.

We record the time to compute triangular decomposition by Algorithm \ref{alg:StrongSfTriDec} (see the column Algorithm \ref{alg:StrongSfTriDec}) and two {\tt Maple} commands (see the column mp-rc) in the group of columns TD.
The two commands used are {\tt RegularChains[Triangularize]} with the options {\tt output=lazard} and {\tt radical=yes}, and  {\tt RegularChains\allowbreak[ChainTools][SeparateSolutions]}. The first one, which is also used in \cite{boulier2014real}, decomposes the system to a finite number of square-free regular chains. 
The second one ensures the zero sets of any two regular chains have no intersection.
In fact, since a strong chain is a regular chain, 
SFSTD is stronger than such decomposition.

In the group of columns RSI, we record the time of real solution isolation computed by the method NRSI (see the column NRSI), the {\tt Mathematica12} command {\tt Solve} (see the column mt-solve) and the {\tt Maple} command  {\tt RootFinding[Isolate]} (see the column mp-rt).

In the group of columns RA, we record the time to compute radicals by the method IRA (see the column IRA) and the {\tt Maple} command {\tt PolynomialIdeals[Radical]} (see the column mp-radical).

\vspace{-3mm}
\subsection{Statistical Experimental Results}

\vspace{-0.5mm}

We show the statistical experimental results of all $151$ examples in Table \ref{tab:result}. 
The number of examples that can be solved within $3600$ seconds is recorded in the row Solved.
The number of LOSS (OT) examples is recorded in the row LOSS (OT).
We record the sum of the computing time of all solved examples (written as solved time) in the row Time (Solved).
For every LOSS or OT example, we record their computing time as $3600$ seconds.
And, the sum of the computing time of all examples is recorded in the row Time.

For triangular decomposition, Algorithm \ref{alg:StrongSfTriDec} performs significantly better than mp-rc. 
There are $31$ examples which can only be solved by Algorithm \ref{alg:StrongSfTriDec}.
And the solved time of Algorithm \ref{alg:StrongSfTriDec} is a half of that of mp-rc.
One main reason why Algorithm \ref{alg:StrongSfTriDec} performs better is that the outputs of Algorithm \ref{alg:StrongSfTriDec} usually have less components.
Denote by $m_1$ and $m_2$ the numbers of components computed by Algorithm \ref{alg:StrongSfTriDec} and mp-rc on the same example, respectively. We observe that $m_1<m_2$ for $61$ examples. Especially, for $54$ of those $61$ examples, we have $m_1\le \frac{1}{2}m_2$. On the contrary, there are no examples where $m_1>m_2.$ And $m_1=m_2$ for $54$ examples where $m_1=m_2=1$ for $39$ examples.



For real solution isolation, mt-solve performs better than NRSI on small examples which can be solved in $2$ seconds, but the solved time of NRSI is approximately $450$ seconds less than that of mt-solve.
And, there are $5$ difficult examples solved successfully by NRSI which cannot be solved by mt-solve.
NRSI solves $3$ examples that mp-rt does not, while mp-rt solves $2$ examples that NRSI does not. 
However, the solved time of NRSI is approximately $10000$ seconds less than that of mp-rt.
It is worth noting that for the system katsura8, the computing time of NRSI is three times that of SFSTD.
This is because the computed SFSTD has huge coefficients. 

To compute radicals, the method IRA solves $145$ examples successfully, while mp-radical solves $140$. The solved time of IRA is about $1200$ seconds less than that of mp-radical.
Since it is difficult to compute intersections of ideals (see the systems redcyc7 and kss3), IRA does not perform as well as we expect.
\begin{table}[t]
\scriptsize
    \centering
    \scalebox{0.8}{
    \begin{tabular}{|c|c||c|c||c|c|c||c|c|}
\hline
\multicolumn{2}{|c||}{}& \multicolumn{2}{c||}{TD}&\multicolumn{3}{c||}{RSI}&\multicolumn{2}{c|}{RA}\\
\hline
sys & {\it n}/{\it d} & Algorithm \ref{alg:StrongSfTriDec}         & mp-rc & NRSI               & mt-solve & mp-rt   &  IRA & mp-radical \\
\hline
nld-4-5&5/4&35.30&OT&35.64&1095.42&1522.17&368.91&88.79\\
nld-6-4&4/6&83.83&OT&86.06&103.09&2918.52&183.19&878.51\\
nld-9-3&3/9&22.22&0.65&47.50&2.71&191.98&39.02&OT\\
nld-10-3&3/10&109.90&0.50&130.42&12.39&670.93&117.18&OT\\
nql-10-4&10/4&0.01&0.09&0.14&0.00&OT&0.02&OT\\
nql-15-2&15/2&0.01&0.17&0.18&0.02&OT&0.01&OT\\
Reif&16/2&0.02&0.80&0.02&OT&0.07&0.02&0.17\\
simple-nql-20-30&20/30&0.01&0.28&0.66&0.12&OT&0.01&LOSS\\
Trinks-2&6/3&0.01&OT&0.02&0.01&0.11&0.03&0.04\\
Trinks-difficult&6/3&0.12&OT&0.16&0.02&0.31&0.18&0.26\\
Uteshev-Bikker&4/3&0.29&OT&0.63&0.25&1.16&0.49&1.22\\

wang\_ex34&14/5&0.09&34.64&0.11&OT&0.21&0.14&0.07\\
wang\_ex40&6/2&0.15&OT&0.77&0.37&0.36&0.26&0.65\\
boon&6/4&0.09&OT&0.19&0.02&0.13&0.13&0.35\\
cpdm5&5/3&19.25&OT&19.64&13.46&2.90&55.57&16.28\\
eco8&8/3&0.33&LOSS&0.75&2.03&0.27&0.44&1.49\\
redcyc6&6/11&1.01&OT&1.34&0.38&1.91&3.38&9.43\\
redcyc7&7/13&44.22&LOSS&55.25&OT&1684.53&570.98&166.93\\

extcyc6&6/6&5.88&OT&7.36&43.90&3.25&38.25&15.62\\

cassou&4/8&0.23&OT&0.29&0.06&0.36&0.35&0.76\\
virasoro&8/2&125.48&OT&128.64&OT&31.66&2477.53&146.24\\
d1&12/3&4.90&OT&8.37&0.77&6.03&13.95&675.68\\
kin1&12/3&22.00&OT&39.90&1.16&19.69&39.74&902.41\\
des18\_3&8/3&22.15&OT&22.83&1.74&1.99&32.89&3.61\\
kinema&9/2&0.85&OT&1.27&1.34&0.56&1.19&2.78\\
rbpl24&9/2&14.54&OT&19.82&6.54&10.87&20.33&408.90\\
reimer5&5/6&0.42&OT&0.97&5.17&1.55&0.60&5.29\\
filter9&9/4&2.66&OT&6.33&4.73&36.24&3.85&45.37\\
katsura6&7/2&0.99&OT&2.92&2.37&0.47&1.35&8.46\\
katsura7&8/2&12.42&LOSS&33.09&40.46&1.68&15.18&29.53\\
katsura8&9/2&291.76&OT&757.49&920.58&13.24&340.98&OT\\
katsura9&10/2&OT&LOSS&OT&OT&151.71&OT&OT\\
katsura10&11/2&OT&OT&OT&OT&2123.81&OT&OT\\
utbikker&4/3&1.25&OT&1.43&0.24&1.31&2.13&3.35\\
kotsireas&6/5&4.16&OT&4.41&2.81&1.59&8.77&6.91\\
chandra6&6/2&1.94&OT&3.27&0.37&0.75&3.03&1.82\\
tangents0&6/2&0.92&OT&1.07&0.10&0.69&1.38&0.76\\
assur44&8/3&4.79&OT&5.95&10.65&2.79&6.33&7.11\\
cyclic6&6/6&1.18&OT&1.52&1.02&1.24&3.93&12.68\\
cyclic7&7/7&76.70&OT&90.17&OT&691.37&859.50&290.58\\
cyclic9&9/9&OT&OT&OT&OT&OT&OT&OT\\
cyclic10&10/10&OT&LOSS&OT&OT&OT&OT&OT\\
cyclic11&11/11&OT&LOSS&OT&OT&OT&OT&OT\\
kss3&10/2&212.92&212.86&216.60&2.67&249.54&OT&540.37\\
\hline

    \end{tabular}}
    \caption{Timings for computing triangular decomposition of zero-dimensional systems, isolating cubes of real solutions, and radicals of zero-dimesional ideals.}
    \label{tab:timing}
\end{table}


\begin{table}[t]
\scriptsize
    \centering
    \vspace{-9mm}
    \scalebox{0.85}{
    \begin{tabular}{|c||c|c||c|c|c||c|c|}
\hline
& \multicolumn{2}{c||}{TD}&\multicolumn{3}{c||}{RSI}&\multicolumn{2}{c|}{RA}\\
\hline
& Algorithm \ref{alg:StrongSfTriDec}         & mp-rc & NRSI               & mt-solve & mp-rt   & IRA & mp-radical \\
\hline
Solved&146&115&146&141&145&145&140\\
\hline
LOSS&0&6&0&0&0&0&0\\
\hline
OT&5&30&5&10&6&6&11\\
\hline
Time (Solved)&1191.37&2842.90&1842.40&2301.10&11238.8&5438.7&6695.10\\
\hline

Time &19191.37&110842.90&19842.40&38301.10&32838.8&27038.7&42695.10\\
\hline
    \end{tabular}}
    \caption{Statistical experimental results of all testing examples (151 examples).}
    \label{tab:result}
    \vspace{-6mm}
\end{table}

\vspace{-3mm}
\section{Conclusion}\label{sec:conclusion}
\vspace{-0.5mm}

In the paper, we propose an algorithm for computing SFSTD and prove that the arithmetic complexity can be single exponential time (note that there are few results about the complexity of triangular-decomposition algorithms).
Our algorithm is partly inspired by \cite[Algorithm 1]{DBLP:conf/casc/DongM19} and thus it is based on Gr\"obner bases.
The novelty of our algorithm is that we make use of separant and saturated ideals to ensure that every strong chain is square-free and the zero sets of any two strong chains have no intersection, respectively.
It is worth noting that although SFSTD is stronger than zs-rc decomposition in \cite{boulier2014real},
our algorithm is much more efficient than the classical method in experiments.
The only disadvantage of our algorithm is that a computed SFSTD of a big system always has huge coefficients.   
So, it sometimes takes a large amount of time to compute isolating cubes of every square-free strong chain or compute intersections of ideals.
We will consider giving a bit complexity analysis of our algorithm in the future.


\vspace{-3mm}
\begin{acks}
This work was supported by the NSFC under grant No. 61732001.
\end{acks}

\vspace{-2mm}
\bibliographystyle{ACM-Reference-Format}
\bibliography{FGT}
\end{document}